\documentclass[graybox]{svmult}

\usepackage{type1cm}        % activate if the above 3 fonts are
\usepackage{makeidx}         % allows index generation

\usepackage{graphicx}        % standard LaTeX graphics tool

\usepackage{multicol}        % used for the two-column index

\usepackage[bottom]{footmisc}% places footnotes at page bottom

\usepackage{newtxtext}       %

\usepackage[varvw]{newtxmath}       % selects Times Roman as basic font

\usepackage{url}

\makeindex             % used for the subject index
\begin{document}

\title*{Design and Application of Energy-saving Sub-Optimal Sliding Mode Control}

% Use \titlerunning{Short Title} for an abbreviated version of
% your contribution title if the original one is too long

\author{Michael Ruderman \orcidID{0000-0002-7021-8713} }

\authorrunning{Energy-saving Sub-Optimal SMC \textcolor[rgb]{0.00,0.00,1.00}{(submitted chapter manuscript)}} % for an abbreviated version of

% your contribution title if the original one is too long

\institute{Michael Ruderman \at University of Agder, P.B. 422, Kristiansand, 4604, Norway, \email{michael.ruderman@uia.no}\\
\textcolor[rgb]{0.00,0.00,1.00}{(Springer submitted chapter manuscript)}}

\maketitle

\abstract{The recently introduced energy-saving extension of the \emph{sub-optimal} sliding mode control (SOSMC), which is known in the literature for the last two and half decades, incorporates a control-off mode that allows for saving energy during the finite-time convergence process. This novel \emph{energy-saving} algorithm (denoted by ES-SOSMC) assumes the systems with relative degree two between the sliding variable and the switching control with a bounded magnitude, while the matched upper-bounded perturbations are not necessarily continuous. The design and practical application of the ES-SOSMC are the subject of this chapter. A method for parameterizing the ES-SOSMC through a constrained minimization of the \emph{energy cost function} is recalled which guarantees the total energy consumption is lower than that of the conventional SOSMC. Also the residual steady-state oscillations (chattering), occurring when additional (actuator) dynamics are taken into account, are addressed. An application example for scanning and machining a rough surface, both of which require a stiff position control in contact with a moving surface, demonstrates practical suitability of the control. Here, ES-SOSMC is compared with SOSMC by showing an equivalent tracking and stabilization performance and evaluating the energy-saving operation with respect to a fuel consumption norm.}

\section{Introduction}
\label{sec:1}

Sliding mode control (SMC) is recognized as one of the most promising robust control techniques, for fundamentals we exemplary refer to the textbooks \cite{utkin1992,edwards1998,shtessel2014,utkin2020}. In discontinuous SMCs, the switching controller aims to force an uncertain system dynamics to evolve into a specified manifold $\sigma=0$ and to stay there for all times $t > T_c$, where $0 < T_c < \infty$ is a finite time of convergence of the state trajectories towards $\sigma=0$. The sliding variable $\sigma(x,t)$, with $x(t) \in \mathbb{R}^n$ to be the vector of the system states, constitutes a continuous function and admits the relative degree which is lower than the relative degree of the system plant itself with the respect to the control channel $u(t)$. Moreover, $\sigma(\cdot)$ function describes the desired motions in the system state-space without perturbations. Therefore, SMC enforces a reduced-order dynamics and can guarantee insensitivity to a certain class of perturbations when staying on the \emph{sliding surface} $\sigma = 0$, i.e., being in the \emph{sliding mode}.  
  
In the second-order SMCs \cite{levant1993,FridmanLevant2002}, also the time derivative
of the sliding variable is forced to zero in a finite time, i.e. $\sigma=\dot{\sigma}=0$, thus allowing for a robust regulation of uncertain systems with the relative degree two between the sliding variable and the control signal. Worth noting is that second-order SMC algorithms belong to a more general class of the higher-order SMC systems, which have been developed over the last two and half decades for robustly controlling uncertain dynamic systems with relative degree two and higher, see e.g. \cite{shtessel2014,utkin2020} and references therein. Applications of higher-order SMC systems can be found in quite various engineering fields. Here to mention just a few of the recent, the higher-order SMCs for e.g. temperature control of chemical fluids in silicon wafer production \cite{koch2020sliding}, states estimation for frequency regulation and economic dispatch in power grids \cite{rinaldi2021}, cascaded regulation of DC–DC boost converters \cite{al2022analysis}, trajectory tracking of unmanned aerial vehicles \cite{rao2024adaptive}, servo-control of hydro-mechanical drives \cite{estrada2025}, and clearly many more others.

Among the practical aspects associated with the second-order SMCs, two most relevant issues can be highlighted here as being essential in multiple applications. First, the time derivative of the sliding variable may be hardy available from the measurements in a system under control. Second, a discontinuous control implies a high (theoretically infinite) switching frequency of the control
variable once in the sliding mode, that can be both hardware-fatiguing and energy-inefficient
for several classes of the system plants. To address the first above-mentioned issue, the so called sub-optimal SMC algorithm (further as SOSMC) was developed \cite{bartolini1997,bartolini2003} based on the bang-bang control principles and the so-called Fuller's problem \cite{fuller1960relay}.

\begin{figure}[!ht]
\sidecaption
\includegraphics[scale=1.1]{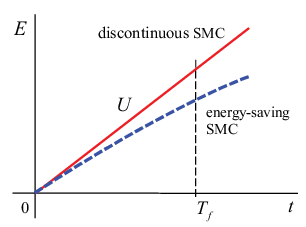}
\caption{Fuel consumption metric (i.e. energy costs) of a continuously switching SMC.}
\label{fig:MR:Fuel}       
\end{figure}

Just as any discontinuous SMC, the SOSMC algorithm executes a continuously commutating
control sequence $u(t) \in \{-U, U\}$, while the frequency of switching between two discrete control states with amplitude $U > 0$ increases towards infinity (theoretically) once the control system is in the sliding mode. Defining the fuel consumption metric and, therefore, the energy costs of a SMC as 
\begin{equation}
E = \int\limits_{0}^{T_f} |u| dt, \label{eq:1:1}
\end{equation}
cf. \cite{athans1964}, it is easy to recognize that for SOSMC (actually  as for any standard discontinuous SMC) the $E$ value grows linearly over the time, i.e. $ E = U \cdot t$, see Fig. \ref{fig:MR:Fuel}. 
Note that the upper integration limit $T_f$ can be interpreted as a final time of the control system operation or, alternately, as a reaching time of certain operation point where the control can be (at least temporary) switched off, thus stopping the associated energy consumption. Therefore, a control approach that would allow for a sequence of time intervals $\in [0, T_f]$ (or at least for one time interval) with 'control-off' i.e. $u=0$ would automatically provide an energy-saving operation, under the condition that the control performance remains the same in terms of e.g. convergence time or output accuracy. Such energy-saving operation mode requires then some fuel-optimized control scheme, cf. the dashed blue line in Fig. \ref{fig:MR:Fuel}; a promising example of that is the novel \emph{energy-saving} sub-optimal SMC algorithm (further ES-SOSMC) which was recently introduced in \cite{ruderman2023energy}. At that point we should recall, however, that a fuel-optimal control is well-known for unperturbed second-order systems \cite{athans1963,athans1964}.

Against the above introductory backgrounds, the purpose and contents of the recent chapter are to provide the reader with consolidated details and features of ES-SOSMC, and to demonstrate its application based on a carefully thought practical example. While an experimental feasibility of the ES-SOSMC implementation and the resulted energy saving were demonstrated on a simple laboratory setup in \cite{ruderman2024}, the application example elaborated in this chapter has larger practical relevance. The addressed application of ES-SOSMC to a scanning and machining process of a moving rough surface address both, the trajectory tracking and point stabilization problems in presence of an application-inbuilt dynamic disturbance with a pre-computable upper bound but without its explicit model. To allow the reader with easy access to the design and use of ES-SOSMC, we purposefully omit many mathematical details and developments in the analysis of ES-SOSMC, while an interested reader can still find them in \cite{ruderman2023energy}. In section \ref{sec:2}, the necessary preliminaries, including the summarized conventional SOSMC are given, while closely following \cite{bartolini2003}. The ES-SOSMC is described in section \ref{sec:3} along with its convergence properties, parameterization which is based on minimizing the energy cost-function, and the describing function-based chattering analysis in case of additional (parasitic) actuator or sensor dynamics. The entire section \ref{sec:4} is dedicated to the elaborated application example, while the chapter is briefly concluded by section \ref{sec:5}.

\section{Preliminaries}
\label{sec:2}

\subsection{Fuel-optimal control problem}
\label{sec:2:sub:1} 

For introducing the ES-SOSMC with its principle 'control-off' extension of SOSMC, it is worth first to recall (at least qualitatively) the fuel-optimal control problem addressed and solved in the earlier works \cite{athans1963,athans1964}. 

\begin{figure}[!ht]
\sidecaption
\includegraphics[scale=1.2]{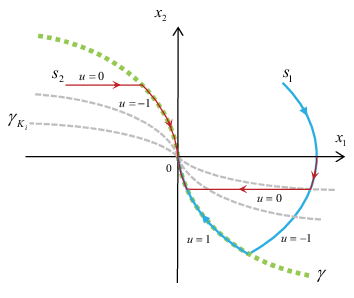}
\caption{Phase-portrait of unperturbed second-order system with time-optimal and fuel-optimal control.}
\label{fig:MR:OptimalPP}       
\end{figure}

Considering a simple unperturbed second-order process $\dot{x}_1 = x_2, \; \dot{x}_2=u$, it is well known (see e.g. seminal literature \cite{Pontryagin1962}) that the time-optimal control solution in terms of the minimum time is given along the 
\begin{equation}\label{eq:2:1:0}
\gamma: \; x_1 + \frac{1}{2} |x_2| x_2 = 0
\end{equation}
trajectory, see Fig. \ref{fig:MR:OptimalPP} (dashed light-green line). The corresponding time-optimal control law is given by
\begin{equation}\label{eq:2:1:1}
u = \left\{
      \begin{array}{ll}
        -1, & \hbox{ if } \; x_1 > -\dfrac{1}{2}|x_2|x_2 \: \vee \: x_1 = -\dfrac{1}{2} x_2^2 \: \wedge \: x_2 > 0, \\[4mm]
        +1, & \hbox{ if } \; x_1 < -\dfrac{1}{2}|x_2|x_2 \: \vee \: x_1 = \dfrac{1}{2} x_2^2 \: \wedge \: x_2 < 0.
      \end{array}
    \right.
\end{equation}
The control \eqref{eq:2:1:1} is also often referred to as a bang-bang control, and it is (intuitively) clear that the minimum time $t^\ast$ achieved in this way depends on the initial conditions $\bigl(x_1(0),x_2(0)\bigr)$, cf. the starting point $s_1$ in Fig. \ref{fig:MR:OptimalPP}, while the control amplitude $U=\max(u)$. The latter, as $\mp U$, will appear in the general case instead of $\mp 1$ in the discrete bang-bang control \eqref{eq:2:1:1}.

Next, the fuel-optimal free response time problem (further with $U=1$ as above for the sake of simplicity) can be formulated as follows \cite{athans1964}:
\\
\emph{For the system of double integrators, with the control values $|u| \leq 1$, find the control $u(t)$ which will drive any initial state $s = \bigl(x_1(0), x_2(0)\bigr)$ to zero equilibrium in a finite time $T_f$ and which will minimize the consumed fuel measured by \eqref{eq:1:1}, while $T_f$ is not specified. Moreover, if more than one control require the same minimal fuel, select such minimal-fuel control which requires also the smallest response time $T_f$.  
}\\
When not limiting $T_f$, it can be shown that the minimal-fuel solution corresponds to the response time $T_f \rightarrow \infty$, cf. \cite{athans1963}. Therefore, the above formulated problem needs to be extended by an (application specific) response time constraint given by 
\begin{equation}\label{eq:2:1:2}
t^\ast \leq T_f \leq K t^\ast,
\end{equation}
where $t^\ast$ is the minimum time, as before, for reaching the origin from a given initial state $s$. Obviously, the prescribed time constraining constant
$$
K > 1
$$
is independent of the initial state $s$. The above constraint should guarantee that the response time $T_f$ does not exceed the prescribed multiple of the minimum time corresponding to each state \cite{athans1964}. Comparing to the time-optimal control solution, which is progressing along the $\gamma$-curve and has the response time $T_f=t^\ast$, the minimal-fuel solution should admit the trajectory segments where $u=0$, cf. Fig. \ref{fig:MR:OptimalPP} (solid dark-red line).
For achieving this goal, in \cite{athans1964} it was shown that the intersection point of the time-optimal trajectory with one of the auxiliary curves
\begin{equation}\label{eq:2:1:3}
\gamma_{K_i} \: : \; x_1 = -\psi |x_2| x_2,
\end{equation}
where $\psi$ is an auxiliary parameter given by 
\begin{equation}\label{eq:2:1:4}
\psi = \dfrac{K}{2K - 1 - 2 \sqrt{K(K-1)}} - \dfrac{1}{2},
\end{equation}
determines the system state where the control must be switched off to $u=0$. The control remains switched-off until the progressing trajectory will cross the $\gamma$-curve given by \eqref{eq:2:1:0}, see Fig. \ref{fig:MR:OptimalPP}. 
It is evident that the chart of $\gamma_{K_i}$ curves with $\psi > 1/2$ represent the set of the fuel-optimal controllers with the specified response time constraint $K_i > 1$. It can also be shown, cf. \cite{athans1964}, that two limit cases for a fuel-optimal control emerge
\begin{equation}\label{eq:2:1:5}
\left. 
  \begin{array}{cc}
   \underset{K_i \rightarrow 1}{\lim} \{ \gamma_{K_i} \} = & \gamma  \\[4mm]
   \underset{K_i \rightarrow \infty}{\lim} \{ \gamma_{K_i} \} = & \{x_1, 0 \}
  \end{array}
\right \}.
\end{equation}
Thus, for any finite time constraining $K_i < \infty$, the state trajectories will always have an interim horizontal segment where $u=0$, cf. both red-colored trajectories from two different starting points $s_1$ and $s_2$ in Fig. \ref{fig:MR:OptimalPP} which imply two different $K_i$ values. 

It becomes (intuitively) clear that once the above considered second-order system is perturbed by some matched disturbance quantity $f(\cdot)$, the trajectory segment where $u=0$ does not necessary remain horizontal. Instead, the disturbance value can force the trajectory to proceed either closer to the time-optimal solution, provided $|f| \leq U$, or to reach the subset $\{x_1, 0\}$ on the ordinate of the phase-plane before attaining the origin, cf. \cite[Figure~2]{ruderman2023energy}. This situation means that the system becomes driven by the disturbance value during a fuel-optimal control is in the 'off'-mode, that precisely motivates introduction of the ES-SOSMC controller \cite{ruderman2023energy}.

\subsection{Sub-optimal sliding mode control}
\label{sec:2:sub:2} 

Considering the class of uncertain dynamic systems for which the well-defined sliding
variable $\sigma(t)$ has the relative degree $r=2$ with respect to the control variable
$u \in \mathbb{R}$, the corresponding dynamics affected by the matched bounded perturbations $f(\cdot,t)$ is written as
\begin{equation}\label{eq:2:2:1}
\ddot{\sigma}(t) = f(\cdot,t) + g(\cdot,t) u(t).
\end{equation}
Here $g > 0$ is the (uncertain) input gain function. Assuming the global boundedness conditions for both sufficiently smooth function
\begin{eqnarray}
| f(\cdot,t) | \leq  \Delta, \label{bndF}\\
\Gamma_m \leq \:  g(\cdot,t) \leq  \Gamma_M, \label{bndG}
\end{eqnarray}
with some known constants $\Delta > 0$ and $0 < \Gamma_m \leq \Gamma_M$, the so-called 
'sub-optimal' second-order SMC algorithm was originally proposed in \cite{bartolini1997}.
The discontinuous SMC feedback law is written, cf. \cite{bartolini2003}, as
\begin{equation}\label{eq:3}
u(t) = \left\{ \begin{array}{cl}
                 -U \mathrm{sign}\bigl(\sigma(t) - \sigma(0)\bigr) & \quad 0 \leq t < t_{M_1}, \\[0.5mm]
                 -\alpha(t) U \mathrm{sign}\bigl(\sigma(t)-\beta_1 \sigma_{M}(t)\bigr) & \quad t \geq t_{M_1},
               \end{array}
  \right.
\end{equation}
where $\beta_1 \in [0;\, 1)$ in the design parameter which affects convergence properties (ratio and shape) of the globally stable equilibrium $\sigma=\dot{\sigma} = 0$. The gaining factor is 
\begin{eqnarray}
\label{eq:3b}
  \alpha(t) &=& \left\{%
\begin{array}{ll}
    1,        & \hbox{ if } \bigl(\sigma(t)-\beta_1 \sigma_{M}(t)\bigr) \sigma_{M}(t) \geq 0
    \\[1mm]
    \alpha^* & \hbox{ if } \bigl(\sigma(t)-\beta_1 \sigma_{M}(t)\bigr) \sigma_{M}(t)<0  \\
\end{array}%
\right.,
\end{eqnarray}
and the piecewise-constant signal $\sigma_{M}(t)$:
\begin{eqnarray}
\label{eq:sigmaMidef}
\hspace{-0.5cm}\sigma_{M}(t)=\sigma(t_{M_i})\triangleq \sigma_{M_{i}}, \;t_{M_i} \leq t < t_{M_{i+1}}, \;i=1,2,\ldots,
\end{eqnarray}
represent the local extrema values. In other words, $t_{M_i}$ is the sequence of time instants at which $\dot \sigma(t_{M_i})=0$ in such a way that $\sigma_{M}(t)$ stores the most recent local maximum, minimum, or horizontal flex point of $\sigma(t)$.

The tuning parameters of the sub-optimal algorithm (i.e. SOSMC) are the control
authority $U > 0$, the 'modulation' factor $\alpha^* \geq
1$, and the 'anticipation' factor $\beta_1$. As shown in the survey paper \cite{bartolini2003}, in order to provide the finite-time stabilization of the system \eqref{eq:2:2:1}-\eqref{bndG}, the tuning parameters must be set in accordance with
\begin{eqnarray}
\label{eq:4}
  U & > & \frac{\Delta}{\Gamma_m},\\[1mm]
  \alpha^* & \in & [1; \infty) \cap \Biggl ( \frac{2\Delta + (1-\beta_1)\Gamma_M U}{(1+\beta_1)\Gamma_m U}; \infty
  \Biggr).
\label{eq:5}
\end{eqnarray}
Note that the condition \eqref{eq:4} sets the minimal control authority which is required to overcome the unknown perturbations and, therefore, to impose the sign of $\ddot{\sigma}$. The conditions \eqref{eq:4} and \eqref{eq:5} ensure the finite-time convergence of $\sigma(t)$ and $\dot \sigma(t)$ to the origin, where the second-order sliding mode appears (afterwards) in its proper sense. In
addition, a stronger inequality than \eqref{eq:5} can be imposed by
\begin{equation}\label{eq:6}
\alpha^* \in  [1; \infty) \cap \Biggl ( \frac{\Delta +
(1-\beta)\Gamma_M U}{\beta_1\Gamma_m U}; \infty \Biggr),
\end{equation}
which is required to guarantee a \emph{monotonic} convergence of
$\sigma(t)$ to zero, cf. \cite{bartolini2003}. That means $\sigma(t)$ will experience at most one zero-crossing during the convergence transient and, then, converge to origin without overshooting. For distinguishing the parametric conditions of SOSMC, the tuning conditions \eqref{eq:4}, \eqref{eq:5} can be referred to as for \emph{twisting convergence}, while the conditions \eqref{eq:4}, \eqref{eq:6} as for \emph{monotonic convergence}, respectively. It is also worth recalling that $\dot{\sigma}(t)$ is not available, so that it is impossible to determine $\sigma_M(t)$ by measuring zero-crossing of $\dot{\sigma}(t)$. Still, $\sigma_M(t)$ can be detected by using only measurements of $\sigma(t)$, for instance using $\lq\lq$Algorithm 2" given in \cite{bartolini1997}. Other zero-crossing detection algorithms are equally possible.

For the sake of simplicity, and in approaching the next introduced energy-saving version of SOSMC, the uncertainty condition \eqref{bndG} is relaxed and a constant unity input gain is assumed, thus leading to $g = \Gamma_m = \Gamma_M = 1$. Note that this can be done by assuming worth case conditions for the actuator input gain, and then a corresponding scaling of the $U$ control authority. Following to that, the inequalities \eqref{eq:4}, \eqref{eq:5} and \eqref{eq:6}
considerably simplify to
\begin{eqnarray}
\label{eq:7}
U & > & \Delta, \\
\label{eq:8}
\beta_1 & > &  \frac{\Delta}{U} \qquad \qquad \hbox{: twisting convergence}, \\
\beta_1 & > &  \frac{\Delta + U}{2U} \qquad \: \hbox{: monotonic
convergence}, \label{eq:9}
\end{eqnarray}
and the modulation function in \eqref{eq:3} is set to $\alpha(t) = 1$.

\section{Energy-saving sub-optimal sliding mode control}
\label{sec:3}

\subsection{Control algorithm}
\label{sec:3:sub:1} 

The introduced in \cite{ruderman2023energy} energy-saving modification of the second-order sub-optimal SMC algorithm (denoted here as ES-SOSMC) is 
\begin{equation}\label{eq:9a}
u(t) = -0.5 U \mathrm{sign}(\sigma - \beta_1 \sigma_M) - 0.5 U
\mathrm{sign}(\sigma - \beta_2 \sigma_M) \quad \forall \; t \in (t_{M_1}; \infty),
\end{equation}
while an additional threshold parameter, comparing to \eqref{eq:3}, is used and complying with the strict condition $\beta_2 < \beta_1$. An initial control action
\begin{equation}\label{eq:9b}
\bar{u}(t) = - U \mathrm{sign} \bigl(\sigma(t) - \sigma(0) \bigr)
\quad \forall \; t \in [0;t_{M_1}]
\end{equation}
is also required to accelerate the achievement of the first extremum at $t_{M_1}$, cf. \cite{bartolini2003}. The meaning of the sequence $t_{M_i}$ as well as the definition of
$\sigma_M(t)$ are the same as in the conventional sub-optimal
algorithm (SOSMC). The modified control law of $u(t)$ consists
of the parallel connection of two distinct conventional sub-optimal controllers, each one with the same control authority $0.5U$ but different anticipation factors
$\beta_1$ and $\beta_2$. Clearly, the energy-saving control phases with zero control correspond to the situation when the sign of $\bigl(\sigma(t) - \beta_1 \sigma_M(t)\bigr)$ and the
sign of $\bigl(\sigma(t) - \beta_2 \sigma_M(t)\bigr)$ are opposite such that the two components of the control \eqref{eq:9a} cancel each other, leading to $u=0$. It is also worth noting that by setting $\beta_2=\beta_1$ one recovers the conventional sub-optimal controller \eqref{eq:3} with $\alpha(t) = 1$.

\subsection{Convergence properties}
\label{sec:3:sub:2} 

The global finite-time convergence of ES-SOSMC \eqref{eq:9a}, \eqref{eq:9b}, meaning   $\sigma(t)=\dot \sigma(t)=0$ holds for all $t \geq T_c$ where $T_c \equiv T_f < \infty$ is the convergence time for a given initial state $\bigl(\sigma(0),\dot{\sigma}(0)\bigr)$, implies 
the parametric conditions
\begin{eqnarray}
   \beta_1 + \beta_2   & > \dfrac{2 \Delta}{U},\label{eq:18} \\
  0  \leq \beta_1     & < \;  1, \label{eq:19} \\
  -1  < \beta_2       & < \; \beta_1
\label{eq:20}
\end{eqnarray}
obtained in the analysis developed in \cite{ruderman2023energy}. 
\begin{figure}[!ht]
\sidecaption
\includegraphics[scale=1.1]{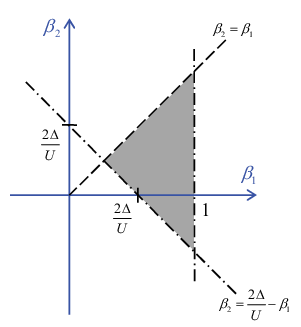}
\caption{Parametric constraints \eqref{eq:18}, \eqref{eq:19},
\eqref{eq:20} of the global finite-time ES-SOSMC convergence in $(\beta_1,\beta_2)$ plane.}
\label{fig:MR:Betas}       
\end{figure}
The imposed $\beta_1, \beta_2$ constraints \eqref{eq:18},
\eqref{eq:19}, \eqref{eq:20} have an illustrative graphical
interpretation shown in Fig. \ref{fig:MR:Betas}.
The $\beta_1, \beta_2$ values admissible for convergence belong to
the grey-shadowed triangle, while on the $\beta_1 = \beta_2$ edge
the energy-saving control \eqref{eq:9a}, \eqref{eq:9b} reduces to
the conventional sub-optimal SMC. Obviously, the $\Delta/U$ ratio
determines the size of an admissible $\{\beta_1, \beta_2\}$ set,
including whether the negative $\beta_2$-values are allowed. It is
also worth noting that the opposite $\beta_2=2\Delta/U - \beta_1$
edge maximizes the $\beta_1-\beta_2$ distance where $u=0$, thus
coming in favor of the energy-saving metric \eqref{eq:1:1}. This can, however,
largely increase the convergence time, including the case $T_c \rightarrow
\infty$, similar as in the case of minimal-fuel control with a free
response time of an unperturbed double-integrator \cite{athans1964}.
At large, the triangle specified by the convergence constraints \eqref{eq:18},
\eqref{eq:19}, \eqref{eq:20} represents the admissible parameterization space of ES-SOSMC, while an energy saving (in comparison to conventional SOSMC) parameterization requires to solve an energy cost-function minimization problem that will be addressed below.

\subsection{Energy cost-function and control parameterization}
\label{sec:3:sub:3} 

The convergence analysis and estimation of the convergence time developed in \cite{ruderman2023energy} based on worst-case scenarios of the state trajectories in the $(\sigma,\dot{\sigma})$-plane reveal that $\sigma(t)=\dot \sigma(t)=0$ for all $t \geq T_c$ where
\begin{equation}\label{eq:61}
T_c \leq \frac{|\dot{\sigma}(0)| }{U-\Delta} + \frac{\sqrt{2 } \, \max \{\Omega_1,  \Omega_2
\} }{1 - \max \{ \sqrt{\eta_1}, \sqrt{\eta_2} \}} \, \sqrt{|\sigma_{M_{1}}|}
\end{equation}
is the upper bound of the convergence time. The constants $\Omega_1, \Omega_2$ are the maximal reaching time \emph{amplification factors} and $0 < \eta_1, \eta_2 < 1$ are the trajectory \emph{contraction factors} between two consecutive extreme values $\sigma_{M_{i}}$ and $\sigma_{M_{i+1}}$ for $i=1,\ldots,\infty$. The subscripts $1$ and $2$ used in both factors distinguish between the boundary cases of the possible state trajectories from $\sigma_{M_{i}}$ to $\sigma_{M_{i+1}}$, see \cite[Figure~2]{ruderman2023energy}, so that the maximum principle is applicable in \eqref{eq:61}. Note that both factors $\Omega_{1,2}$ and $\eta_{1,2}$ depend solely of the $(U, \Delta, \beta_1, \beta_2)$ parameters of the control system, for exact calculations see  \cite{ruderman2023energy}. At that point, it is also worth noting that both the ES-SOSMC and its recovery to SOSMC when $\beta_2=\beta_1$ execute an infinite sequence of the local extrema during the convergence. However, the corresponding infinite sum of the worst-case reaching times between two consecutive $\sigma_{M_{i}}$ and $\sigma_{M_{i+1}}$ extrema develop into a geometric series with
\begin{equation}\label{eq:62}
\sum \limits_{i=1}^{\infty} \sqrt{|\sigma_{M_{i}}|} =
|\sigma_{M_{1}}|^{\frac{1}{2}} \sum \limits_{n=0}^{\infty}
\eta^{\frac{1}{2}n} =  \dfrac{\sqrt{|\sigma_{M_{1}}|}
}{1-\sqrt{\eta}},
\end{equation}
which is convergent. The obtained this way upper bound of the convergence time allows to exclude all 'control-off' phases with $u=0$ and then determine similarly the reaching amplification factors $\tilde{\Omega}_1, \tilde{\Omega}_2$ for the 'control-on' phases only, i.e. when $u\neq0$.  Subsequently, excluding the initial reaching phase (see \eqref{eq:61}), the energy cost functions were determined to represent fuel consumption during convergence \cite{ruderman2023energy}, for the ES-SOSMC as 
\begin{equation}\label{eq:65}
J(\beta_1, \beta_2, U, \Delta) = \frac{ \max \{\tilde{\Omega}_1,
\tilde{\Omega}_2 \} }{1 - \max \{ \sqrt{\eta_1}, \sqrt{\eta_2} \}
}
\end{equation}
and for its conventional SOSMC counterpart as
\begin{equation}\label{eq:66}
\hat{J}(\beta_1, U, \Delta) = \frac{ \hat{\Omega} }{1 - \sqrt{\hat{\eta}} }.
\end{equation}
In the latter and further below, the circumflex notation of the same variables and parameters is used to denote the SOSMC-related quantities.

\begin{figure}[!ht]
\sidecaption
\includegraphics[scale=0.5]{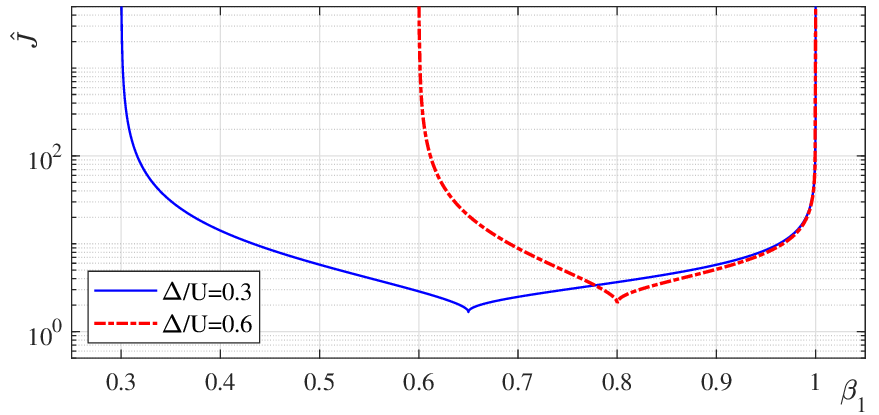}
\caption{Exemplary convergence cost function $\hat{J}$ of SOSMC in dependency of $\beta_1$.}
\label{fig:MR:Beta1}       
\end{figure}
In order to find an energy-saving pair of the switching control threshold values  $(\beta_1,\beta_2)$, namely for some fixed control process parameters $U>\Delta$, the following minimization problem \cite{ruderman2023energy}
\begin{equation}\label{eq:67}
\underset{\beta_1,\beta_2} {\min} \, \Bigl[  J(\beta_1, \beta_2,
U, \Delta) - \hat{J}(\beta_1, U, \Delta) \Bigr]
\end{equation}
is formulated under the hard constraint
\begin{equation}\label{eq:68}
J(\beta_1, \beta_2, U, \Delta) - \hat{J}(\beta_1, U, \Delta) < 0.
\end{equation}
\begin{figure}[!ht]
\sidecaption
\includegraphics[scale=.55]{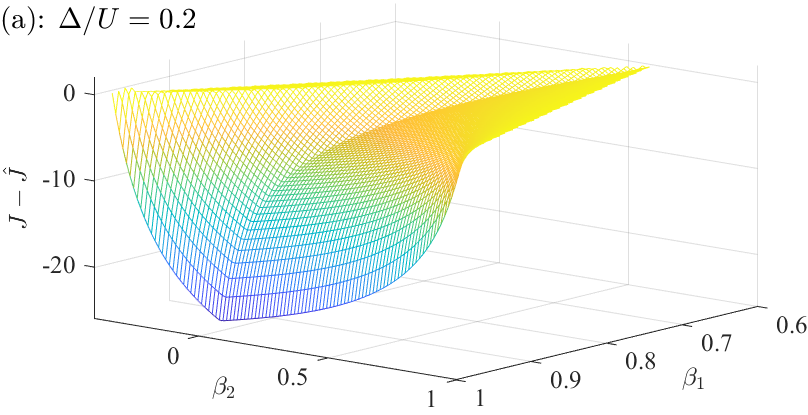}\\
\includegraphics[scale=.55]{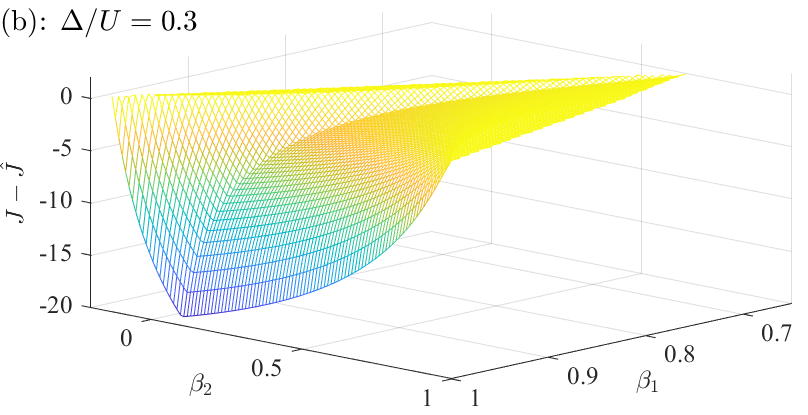}\\
\includegraphics[scale=.55]{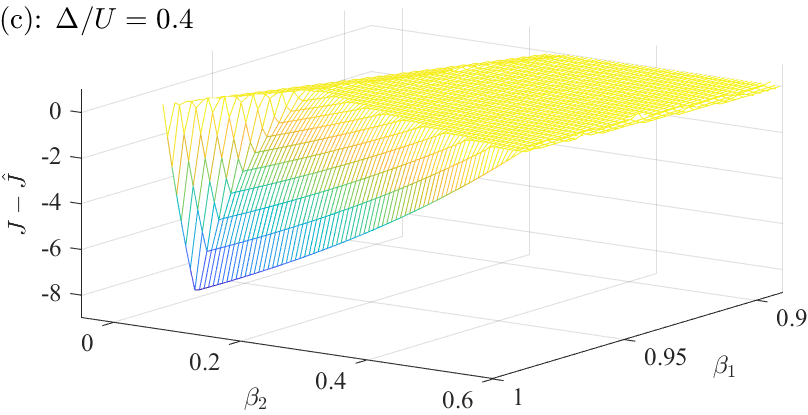}
\caption{Constrained objective function $(J-\hat{J})$ of the
$\beta_1$, $\beta_2$ parameters of ES-SOSMC for the disturbance-to-control ratios: 
$\Delta/U=0.2$ in (a), $\Delta/U=0.3$ in (b), $\Delta/U=0.4$ in (c).}
\label{fig:MR:Optim}       
\end{figure}
The latter condition is strictly required in order to guarantee that for any fixed $\beta_1$ the ES-SOSMC consumes less energy in comparison to the conventional SOSMC. Furthermore, un upper bound 
\begin{equation}\label{eq:69}
\hat{J}(\beta_1, U, \Delta) < \hat{J}_{\max}
\end{equation}
must be set as an additional hard constraint, in order to avoid $\beta_1 \rightarrow 1$ or $\beta_1 \rightarrow \Delta/U$ that would lead to $\hat{T}_c \rightarrow \infty$. This situation is exemplarily computed by \eqref{eq:66} for two different disturbance-to-control ratios and visualized in Fig. \ref{fig:MR:Beta1}, thus illustrating the necessity of introducing $\hat{J}_{\max}$. This way, the SOSMC convergence time $\hat{T}_c$ becomes upper bounded in the optimization problem \eqref{eq:67}. That means for any fixed $\beta_1$ value complying with \eqref{eq:69}, there is an optimal $\beta_2$ counterpart as solution of \eqref{eq:67}. It should be noted, however, that the formulated optimization problem \eqref{eq:67} does not take explicitly into account those transient phases where the control is off, i.e. $u=0$. Though, this does not imply any issues since the overall finite time convergence of ES-SOSMC is upper bounded by \eqref{eq:61}. The numerically solved results of the above formulated constrained optimization problem are visualized in Fig. \ref{fig:MR:Optim} for the $(\beta_1,\beta_2)$ parameters, assuming exemplary the  disturbance-to-control ratios $\Delta/U=\{0.2, \, 0.3, \, 0.4 \}$, see diagrams (a), (b) and (c) respectively.

\subsection{Parasitic actuator dynamics and chattering}
\label{sec:3:sub:4}

If the system under control is subject to an additional
actuator or sensor dynamics, the second-order sliding mode
experiences the residual steady-state oscillations, known as
\emph{chattering}, see e.g. \cite{bartolini1998,boiko2007}. Here it is worth noting that 
chattering, as a high-frequency and rather parasitic side-effect in real sliding modes, appears to be inherently unavoidable and can only be mitigated to a certain degree, cf. \cite{LeeUtkin2007,utkin2015}.

\begin{figure}[!ht]
\sidecaption
\includegraphics[scale=0.9]{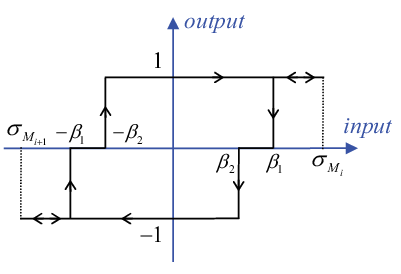}
\caption{Hysteresis relay representation of ES-SOSMC at residual steady-state oscillations due to additional actuator/sensor dynamics.}
\label{fig:MR:Relay}       
\end{figure}

For the first-order actuator (and/or sensor) dynamics with a time constant $\mu > 0$,
the control variable $u(t)$ in \eqref{eq:2:2:1} must be substituted by
a new control variable $v(t)$, which is the solution of
\begin{equation}\label{eq:72}
\mu \dot{v}(t) + v(t) = u(t).
\end{equation}
Assume (based on the previous studies of chattering in second-order SMC systems) that the stable residual steady-state oscillations of the system \eqref{eq:2:2:1}, with the new input channel $v$ and control \eqref{eq:9a}, \eqref{eq:9b} are established. Then, the input-output map of ES-SOSMC takes the form of a three-state hysteresis relay \cite{ruderman2023energy} as shown in Fig. \ref{fig:MR:Relay}.
Note that at steady-state oscillations, the switching thresholds keep the
constant values and are symmetric with respect to zero, so that
the amplitude of oscillations is determined by a cyclic extreme
value $|\sigma_{M_i}|=|\sigma_{M_{i+1}}|$ denoted further by $\sigma_A$. Also we note that if
$\beta_2=\beta_1$, the above input-output map reduces to a standard two-state hysteresis relay, used for harmonic balance analysis of SOSMC \cite[chapter~5.2]{boiko2009}.

\begin{figure}[!ht]
\sidecaption
\includegraphics[scale=1.2]{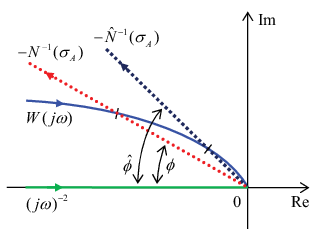}
\caption{Negative reciprocal of DF of the conventional and energy-saving sub-optimal SMC crossing the Nyquist plot of $W(j\omega)$.}
\label{fig:MR:DFnyquist}       
\end{figure}

Based on the describing function (DF) analysis, see e.g. \cite{Atherton75} for fundamentals,
the harmonic balance equation  
\begin{equation}\label{eq:73}
N(\sigma_A) \, W(j\omega) + 1 = 0,
\end{equation}
with DF of the corresponding SMC denoted by $N(\sigma_A)$ and the transfer function of the double integrator augmented by the actuator dynamics \eqref{eq:72} denoted by $W(\cdot)$, must have a real solution in terms of the amplitude $\sigma_A$ and angular frequency $\omega$ for chattering to occur, cf. \cite{LeeUtkin2007,boiko2009}. While the transfer function of the linear part of the closed loop control system results in
\begin{equation}\label{eq:74}
W(j\omega) = \frac{1}{s^2(\mu s + 1)} \Biggr|_{s=j\omega} =
\frac{-1}{\omega^2 + j \cdot \omega^3 \mu},
\end{equation}
the DF of the conventional SOSMC is given by, see \cite{boiko2006,boiko2009},
\begin{equation}\label{eq:75}
\hat{N}(\sigma_A) = \frac{4U}{\pi \sigma_A}
\Bigl(\sqrt{1-\beta_1^2} + j \beta_1 \Bigr).
\end{equation}
We recall that the graphical interpretation of the harmonic balance \eqref{eq:73} solution is 
the intersection of the negative reciprocal of DF with the Nyquist plot of \eqref{eq:74}, cf. Fig. \ref{fig:MR:DFnyquist}. 
The negative reciprocal of \eqref{eq:75} is a straight line lying in the 2nd quadrant and inclined by the angle, cf. \cite{boiko2009}, 
\begin{equation}\label{eq:76}
\hat{\phi} = \arctan \Biggl(-\frac{\beta_1}{\sqrt{1-\beta_1^2}}.
\Biggr)
\end{equation}
It can be recognized that larger $\beta_1$ values lead to a larger $|\hat{\phi}|$ and, thus, shift the locus intersection point towards higher frequencies and lower amplitudes.

Since ES-SOSMC constitutes a linear combination of two negative hysteresis relays, its DF is obtained 
as a summation of two corresponding DFs leading to, \cite{ruderman2023energy},
\begin{equation}\label{eq:76}
N(\sigma_A) = \frac{2U}{\pi \sigma_A} \Bigl(\sqrt{1-\beta_1^2} +
\sqrt{1-\beta_2^2} + j (\beta_1 + \beta_2) \Bigr).
\end{equation}
From evaluating the negative reciprocal of \eqref{eq:76}, the angle argument is 
\begin{equation}\label{eq:77}
\phi = \arctan
\Biggl(-\frac{\beta_1+\beta_2}{\sqrt{1-\beta_1^2}+\sqrt{1-\beta_2^2}}
\Biggr).
\end{equation}
One can recognize that for a fixed $\beta_1$ value, the ES-SOSMC has always a lower angle $|\phi|$ 
than the SOSMC has, while for $(\beta_1-\beta_2) \rightarrow 0$ the angle $\phi \rightarrow \hat{\phi}$, cf. Fig. \ref{fig:MR:DFnyquist}. Following to that, can directly conclude that the corresponding steady-state oscillations have a higher amplitude and a lower frequency for ES-SOSMC compared to those of the conventional SOSMC. It is also worth noting that unlike the solely double integrator (depicted by the horizontal green line in Fig. \ref{fig:MR:DFnyquist}), the transfer characteristics  $W(j\omega)$ have the Nyquist plot which is lying also in the 2nd quadrant at higher angular frequencies. Thus, it has always an intersection point with the negative reciprocals of DF. This implies an existence of $(\sigma_A, \omega)$ solutions of the harmonic balance equation \eqref{eq:73} and therefore chattering. From the solution of the harmonic balance equation \eqref{eq:73} with \eqref{eq:74} and \eqref{eq:76}, the chattering parameters of the ES-SOSMC were obtained in \cite{ruderman2023energy} as the oscillations angular frequency
\begin{equation}\label{eq:78}
\omega_c = \mu^{-1} \frac{\beta_1+\beta_2}{
\sqrt{1-\beta_1^2}+\sqrt{1-\beta_2^2}}
\end{equation}
and the corresponding oscillations amplitude 
\begin{equation}\label{eq:79}
\sigma_A = \frac{\sqrt{\mu^2 \omega_c^2 + 1}}{\omega_c^2 \bigl(
\mu^2 \omega_c^2 + 1 \bigr) }.
\end{equation}

\section{Application example}
\label{sec:4} 

\begin{figure}[!ht]
\sidecaption
\includegraphics[scale=.5]{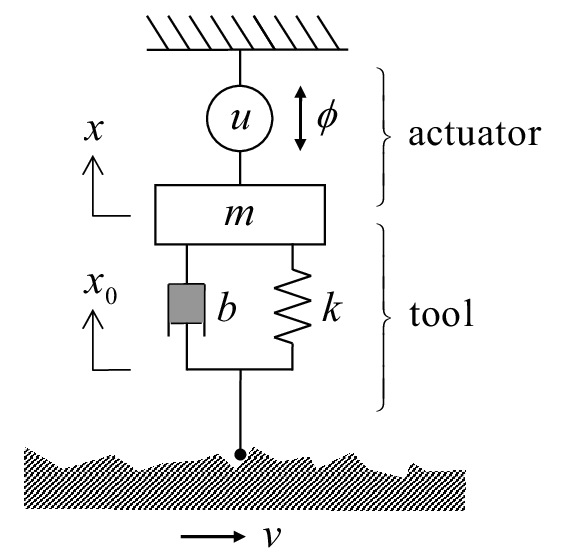}
\caption{Mechanical model of the actuator and tool in contact with a moving rough surface.}
\label{fig:MR:application}       
\end{figure}

Stiff position control in contact with a moving surface is a meaningful and promising application for ES-SOSMC, where a constant reference value in either relative or absolute displacement coordinates must be kept despite a kinematic excitation by a randomly rough surface. A typical mechanical structure of the actuator with tool in contact with the moving surface is shown in Fig. \ref{fig:MR:application}. The stiffness and damping of the mechanical tool are parameterized by $k$ and $b$, respectively. Since the actuator and the tool are rigidly connected and have the same degree of freedom $x$, their total (lumped) mass is summarized in $m$. Over and beyond, for the same reason of a rigid actuator-tool assembly  the damping parameter $b$ may include also an inherent damping of the actuator itself, i.e. its linear part. The tool tip slides over an irregular surface moving in the orthogonal direction to $x$ with a (constant) velocity $\nu$, which is a process parameter. The surface asperities appear as excitation sources in the relative coordinates $x_0$, while a basic assumption of non-separation between the tool and the moving surface is made. Without any particular limitation to the actuator type, the produced actuator force in the generalized coordinates is denoted by $u$, while the actuator disturbances are captured by $\phi$ and  bounded by $|\phi| < \Phi < \max (u)$. The actuator disturbances are not necessarily known (or they are mostly weakly known) and can include, for instance, the force ripples due to structural properties of an electro-magnetic actuator, additional viscous and structural damping in the linear-displacement actuators (e.g. magnetic, pneumatic, and hydraulic), or nonlinear friction in any type of actuators \cite{ruderman2023} with mechanical bearings.   

Based on the general mechanical structure specified above, the dynamic system behavior is described by
\begin{equation}
m \ddot{x} = k(x_0 - x) + b(\dot{x}_0 - \dot{x}) + u + \phi. \label{eq:4:1}
\end{equation}
For the physically reasonable roughness and velocity of the surface and a bounded actuator force, 
an upper bound of the external force excitation $F>0$ can be assumed without loss of generality, i.e.
\begin{equation}
\bigl| k(x_0 - x) + b(\dot{x}_0 - \dot{x}) \bigr|  \leq  F < \infty. \label{eq:4:2}
\end{equation}
One can recognize that the dynamic system \eqref{eq:4:1} with \eqref{eq:4:2} corresponds to the perturbed double-integrator \eqref{eq:2:2:1} while
\begin{equation}
\Delta = \frac{F+\Phi}{m}. \label{eq:4:3}
\end{equation}
Here we also recall that only the directly assigned sliding variable $\sigma \equiv x$  must be available from sensing when used for control, eventually also $x_0$ depending on the application (see section \ref{sec:4:sub:2}) but not its time derivative $\dot{\sigma} = \dot{x}$. 

The control task for the system plant \eqref{eq:4:1} is then specified in terms of a robust stabilization of the double integrator \eqref{eq:2:2:1} with \eqref{eq:4:3}, in presence of both internal actuator disturbances and external kinematic excitation by the moving rough surface. Below, two representative application scenarios of such a stiff (i.e. with a relatively high bandwidth) position control are provided in sections \ref{sec:4:sub:2} and \ref{sec:4:sub:3} for the reference values $(x_0-x)_{\textmd{ref}} = \textrm{const}$ and $x_{\textmd{ref}} = 0$, respectively. It is also worth noting that the assumed mechanical structure shown in Fig. \ref{fig:MR:application} is very similar to the well-known modeling of active and semi-active suspension systems, cf. \cite{tseng2015}. Despite such control-oriented modeling is generally suitable for industrial applications of active suspensions in the ground vehicles, see e.g. \cite{tavares2020}, we are not considering it because of missing the double integrator structure for a suspension system under control. Recall that most of the suspension control problems are formulated in terms of minimizing $|\ddot{x}|$ and not $|x|$ in response to the excitation by a moving surface, cf. \cite{tseng2015}.

\subsection{Randomly rough surface}
\label{sec:4:sub:1}

For describing an interaction between the mechanical tool and irregular surface, cf. Fig. \ref{fig:MR:application}, and thus the kinematic excitation provided by the latter, a randomly rough surface is assumed. For theoretical backgrounds of rough surfaces to be modeled as a random process we refer to \cite{nayak1971}, while an isotropic Gaussian surface can be assumed for $x_0$ without loss of generality. Worth noting is that this is well in accord also with Persson's contact theory, cf. \cite{persson2006}, where the height distribution of single asperities is Gaussian, also with a roll-off region in power spectra of the surface roughness. More precisely, a roughness surface is represented as a stationary Gaussian stochastic process. The power spectral density (PSD) of $x_0$ is then given by
$$
S_{PSD}(v) = R v^{\, \beta},
$$
where $v$ is the spatial frequency and $R$ is the roughness coefficient of the surface. The
exponent factor is often approximated by $\beta = -2$.  When the surface is assumed to move with a constant relative velocity $\nu$ (i.e. relative to the horizontal position of the tool as in Fig. \ref{fig:MR:application}), the temporal excitation frequency $\omega$ and the spatial excitation frequency $v$ are related to each other by $\omega = 2 \pi v \nu$.
Then, with respect to $S(\omega) d\omega = S(v) dv$, the displacement power spectral density as a function of the temporal frequency can be written as
\begin{equation}
S_{PSD}(\omega) = \frac{2 \pi R \nu}{\omega^2}. \label{eq:4:4}
\end{equation}
Note that the given above displacement power spectral density $S_{PSD}(v)$, and so $S_{PSD}(\omega)$, is not valid for very low frequencies. Thus, in order to limit displacement to be finite at vanishingly small spectral frequencies, the cutoff $v_0$ cycle/m is used so that from \eqref{eq:4:4} we obtain  
\begin{equation}
S_{PSD}(\omega) = \frac{2 \pi R \nu}{\omega^2 + \omega_0^2}, \label{eq:4:5}
\end{equation}
where $\omega_0 = 2 \pi v_0 \nu$ appears as a process constant for $\nu = \textmd{const} > 0$. 
Based thereupon, the excitation of the vertical disturbance displacement $x_0(t)$ can be described by the first-order filtered transfer characteristics
\begin{equation}
x_0(s) = G(s) w(s) =  \frac{\sqrt{ 2 \pi R \nu }}{s + \omega_0} \, w(s) \label{eq:4:6}
\end{equation}
expressed in Laplace domain with the complex variable $s$, while $w(s)$ is the white noise signal with an appropriately limited power level. Note that such modeling of excitation by the surface roughness becomes widely accepted and standard in the vehicular suspension technologies, cf. e.g. \cite{hrovat1997survey,zuo2003}. Moreover, the power spectral density specifying the surface roughness and disclosing the first-order low-pass filtered characteristics, hence with roll-off region in power spectra similar to \eqref{eq:4:6}, can also be found for microscopic ranges, cf. e.g. \cite{elson1995,gong2016surface}.

\subsection{Scanning of moving rough surface}
\label{sec:4:sub:2}

The main application scenario that we discuss requires the control system to keep a constant distance between a rough surface which is moving horizontally and the actuator centre of mass, cf. Fig. \ref{fig:MR:application}. This implies $x(t) \rightarrow x_{\textmd{ref}}(t) = x_0(t) + X$, meaning the actuator is 'scanning' over the surface with a certain constant distance to it i.e. $X = \textmd{const}  \neq 0$, while the latter is specified by an application. Such scenario is typical for atomic force microscopy (AFM), see e.g. \cite{yong2012invited}, where a flexible cantilever is in contact (via an ultra-sharp tip) with the moving sample surface, and the vertical feedback controller must ensure that a specified cantilever deflection (equivalent to $X$) and so an interaction force remain possibly constant during the scanning process. 

\begin{figure}[!ht]
\sidecaption
\includegraphics[scale=.6]{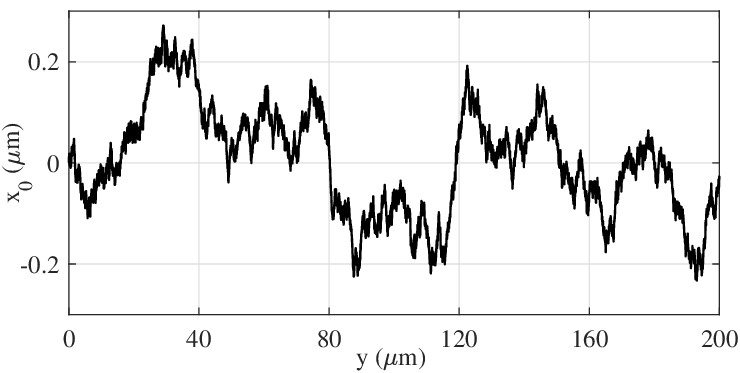}
\caption{Generated surface roughness profile $x_0$ over the horizontal displacement $y$.}
\label{fig:MR:surfacescan}       
\end{figure}

While a typical atomic force microscope  scanning speed is about few tens of microns ($\mu m$) per second, cf. e.g. \cite{sulchek2002}, we assume a scanning speed and thus a constant horizontal relative velocity $\nu = 100 \; \mu m /s$  as a realistic process parameter of the modern AFMs, cf. \cite{dai2018fast}. Worth emphasizing is that the maximum scan speed of an AFM is limited by the
bandwidth of its vertical feedback control loop, see \cite{yong2012invited}. Further we assume without loss of generality (see the measured vertical profiles in e.g. \cite{dai2018fast,wang2021advancing}) that the surface roughness allows for maximal peak-to-peak about 0.5 $\mu m$. The latter, as $\max(x_0)$, allows to assign a reasonable $x_0(s)$ profile according to \eqref{eq:4:6}. The simulated rough surface profile $x_0$ over the horizontal displacement
$$
y = \int \nu dt
$$
is shown for the sake of illustration in Fig. \ref{fig:MR:surfacescan}, for a constant speed $\nu = 100 \; \mu m /s$.

For approaching a real AMF system, the main system parameters are assigned similar to those from the AFM studies \cite{cumpson2004,yong2012design}. The overall lumped mass of a MEMS (micro-electro-mechanical systems) actuator and cantilever tool is assigned to $m = 0.0005$ \emph{kg}, cf. \cite{yong2012design}. A typical cantilever stiffness $k = 0.73$ \emph{N/m} is assumed, cf. \cite{cumpson2004},  thus leading to the natural frequency $\omega_0 = \sqrt{k/m} = 38.21$ \emph{rad/s} of the actuator with tool. Note that the natural frequency value should not be confused with a high fundamental resonance frequency of the cantilever itself, which is usually in kHz range, see \cite{cumpson2004}. The simplest and most popular approach to suppress such high-frequency resonance peaks is to use the notch filters or inversion-based filters in the loop, cf. \cite{yong2012invited}. Such a feed-forward (filtering) compensator reshapes the control signal $u$ at relatively high frequencies only, so that its impact can be largely neglected from the robust SMC viewpoint. Further we note that alternatively to the piezoelectric stack actuators with hight resonance frequencies (typically in kHz range), that is owing to the flexure design with a very high stiffness cf. \cite{yong2012design,yong2012invited}, the voice coil actuators (i.e. Lorentz actuators) can also be used in AFM, cf. \cite{ito2018}, since having a low stiffness between the mover and the base. Over and beyond, a relatively low lumped damping value of $0.0001$ \emph{Ns/m} is assumed for a fixed actuator-tool pair, while the linear damping is mainly due to the structural dissipation  in both the cantilever and (often used) flexure-based nanopositioning stages. The pre-calculated upper bound of the surface-induced and also actuator-inherent disturbances is assigned as $F+\Phi = 0.00003$ \emph{N}, while $\phi$ is mainly due to the nonlinear damping effects with hysteresis, like for example internal friction, cf. \cite{ruderman2023}. Note that, at the same time, $\phi$ does not include the gravity of the entire moving mass, since this can be pre-compensated directly in a feed-forward manner the known constant value.

\begin{figure}[!ht]
\sidecaption
\includegraphics[scale=.6]{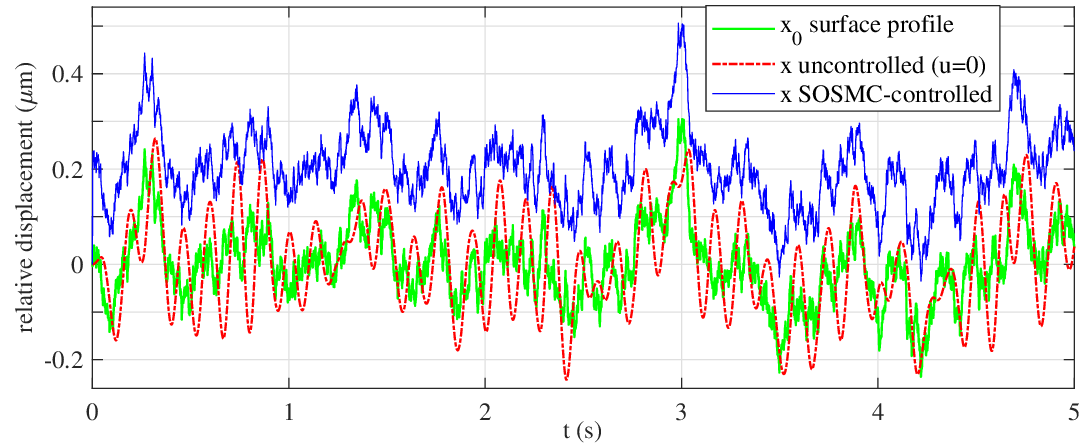}
\caption{Vertical relative displacement: uncontrolled $x(t)$ excited by the rough surface $x_0(t)$ versus the SOSMC-controlled $x(t)$ with $X=2 \: \mu m$ reference.}
\label{fig:MR:noncontrolled}       
\end{figure}

The simulated surface profile $x_0(t)$ and the excited relative displacement of the actuator $x(t)$ are shown one above the other in Fig. \ref{fig:MR:noncontrolled}. For the sake of comparison, the SOSMC-regulated actuator displacement $x(t)$ is also shown for the assigned reference distance $X=0.2 \: \mu m$. The SOSMC is parameterized by $\beta_1 = 0.65$ and $U = 0.2$ \emph{N}, while the latter corresponds to the $\Delta/U = 0.3$ ratio, cf. Fig. \ref{fig:MR:Beta1}. 

\begin{figure}[!ht]
\sidecaption
\includegraphics[scale=.6]{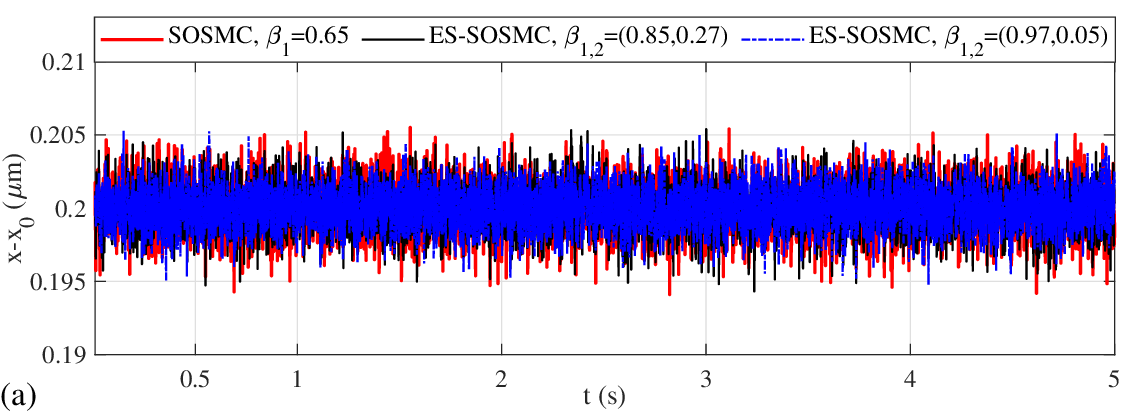}
\includegraphics[scale=.6]{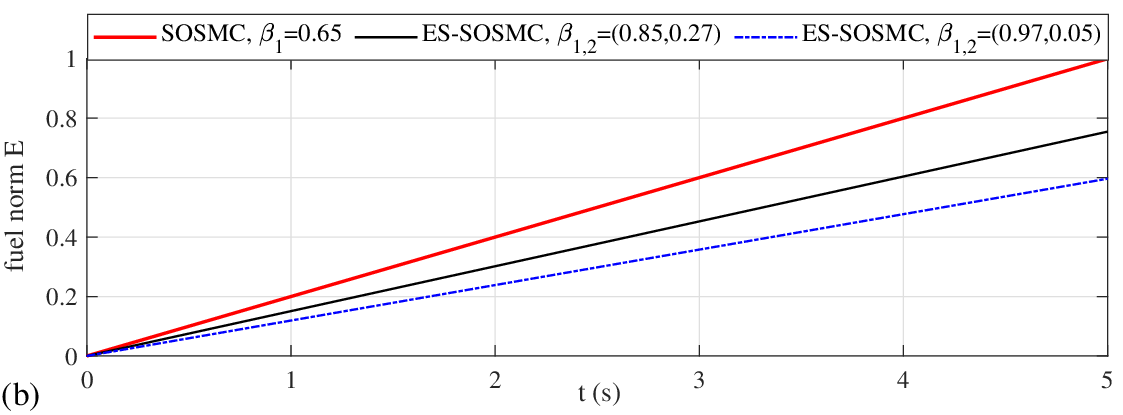}
\caption{(a) target relative distance $x(t)-x_0(t)$; (b) corresponding fuel consumption norm $E$.}
\label{fig:MR:controlled1}       
\end{figure}

Based on the solved constrained optimization problem, for the disturbance ratio $\Delta/U = 0.3$ cf. section \ref{sec:3:sub:3}, two pairs of the locally optimal parameters $(\beta_1, \beta_2)_{i} = (0.85,0.27)$ and $(\beta_1, \beta_2)_{ii} = (0.97,0.05)$ are assigned for ES-SOSMC. For a fair comparison, the globally optimal $\beta_1 =0.65$ assignment is done for the conventional SOSMC, this with respect to the energy cost function also computed for $\Delta/U = 0.3$, cf. Fig. \ref{fig:MR:Beta1}. The control results are compared to each other in Fig. \ref{fig:MR:controlled1} (a). The shown relative target distance $x(t)-x_0(t)$, which should be kept as constant as possible and approaches $X=0.2 \: \mu m$, discloses a fairly similar pattern for all three control configurations, cf. also with Fig. \ref{fig:MR:noncontrolled}. This indicates that ES-SOSMC is applicable here without decreasing the scanning control performance in comparison to SOSMC.  
The energy saving performance is visible from the plot (b), where the fuel consumption norm $E$, according to \eqref{eq:1:1}, is shown over the time. Recall that while an energy-optimal (with respect to $\Delta/U$) assignment of $\beta_1$ for the SOSMC provides the upper bound $E_{sosmc} = U t$ of the fuel consumption, any $E(t)$ curve lying below $E_{sosmc}$ will reveal an energy-saving control behavior. Both $(\beta_1,\beta_2)$ configurations of the ES-SOSMC feature it, while the fuel saving (and thus the energy saving) increases continuously over the time for a persistent disturbance of the scanning process. Also note that both $(\beta_1,\beta_2)_{i}$ and  $(\beta_1,\beta_2)_{ii}$ constitute the computed local energy-saving optima for the assumed $\Delta/U$ ratio, cf. Fig. \ref{fig:MR:Optim} (b), when fixing the $\beta_1$ value by some application requirements, for instance regarding the convergence time upper bound, cf. \eqref{eq:61}, or admissible steady-state chattering of the output (position) variable, cf. \eqref{eq:78}, \eqref{eq:79}.

\subsection{Machining of moving rough surface}
\label{sec:4:sub:3}

Another typical application scenario concerns the machining of a moving surface, which requires $x(t) \rightarrow x_{\textmd{ref}} = \textmd{const}$ despite a continuously acting unknown and highly dynamical disturbance by $x_0(t)$. Such control requirements can be found in e.g. various material cutting and grinding processes, cf. e.g. \cite{altintas2004chatter,aurich2009modelling,brand2025active}. Also, it can be found in the surface  milling processes, cf. e.g. \cite{liang2004machining} and references therein.

When machining a moving rough surface, the tool stiffness $k$ increases significantly more than the increase in the tool and actuator mass $m$ comparing to a scanning process addressed above. This results in a principally higher natural frequency $\omega_0$ comparing to the previous application case. The actuator bandwidth must be respectively higher than $\omega_0$, while the control amplitude $U$ will be correspondingly dimensioned with respect to the pre-estimated $\Delta$ bound of the inherent surface-induced disturbance. At the same time, a trajectory tracking problem addressed in section \ref{sec:4:sub:3} reduces then to the stabilization problem for a fixed equilibrium point $(x_{\textmd{ref}}, 0)$. Due to a thus reduced control problem, from perspective of the system dynamics, and scalability of the control and plant parameters (compared to section \ref{sec:4:sub:3}), a similar design procedure and similar energy saving performance are achieved. For this reason, a detailed presentation is omitted here for the sake of brevity.

\section{Conclusions}
\label{sec:5}

In this chapter, a novel energy-saving sub-optimal sliding-mode control (ES-SOSMC), which was introduced in \cite{ruderman2023energy} as an extension of the well-known sub-optimal sliding-mode control (SOSMC) \cite{bartolini1997,bartolini2003}, has been summarized and shown for use in a suited application scenario. Belonging to the second-order SMC algorithms, for which $\sigma=\dot{\sigma}=0$ is enforced despite the upper-bounded matched perturbations, the ES-SOSMC was first motivated by recalling the fuel-optimal control problem for double integral plants with the response time constraints \cite{athans1963,athans1964}. Then, the ES-SOSMC approach was given along with the convergence properties and optimization solution of the energy cost-function, which allows to determine the control parameters. The previously developed analysis and presentation, cf. Fig. \ref{fig:MR:Optim}, disclose an efficiency of the ES-SOSMC approach for disturbance-to-control ratios $\Delta/U < 0.35$. In order to address the unavoidable parasitic chattering, that appears in real sliding modes due to additional actuator or sensor dynamics in the control loop, the estimates of amplitude and frequency of the self-induced oscillations (i.e. chattering) were also shown based on the harmonic balance analysis developed for ES-SOSMC in \cite{ruderman2023energy}. The second part of the chapter provided the detailed description of a possible application scenario of scanning and/or machining a moving rough surface. Here the robust ES-SOSMC allows continuously an energy-saving operation while preserving the same tracking and stabilization performance in comparison with the standard SOSMC, which was assumed as a benchmarking second-order SMC system.

%\begin{acknowledgement}
%If you want to include acknowledgments of assistance and the like at the end of an individual %chapter please use the \verb|acknowledgement| environment -- it will automatically be rendered in %line with the preferred layout.
%\end{acknowledgement}

\bibliographystyle{plain}        % Include this if you use bibtex
\bibliography{references}

\begin{thebibliography}{10}

\bibitem{al2022analysis}
Ayman Al~Zawaideh and Igor Boiko.
\newblock Analysis of stability and performance of a cascaded {PI} sliding-mode
  control {DC}--{DC} boost converter via {LPRS}.
\newblock {\em IEEE Transactions on Power Electronics}, 37(9):10455--10465,
  2022.

\bibitem{altintas2004chatter}
Yusuf Altintas and Manfred Weck.
\newblock Chatter stability of metal cutting and grinding.
\newblock {\em CIRP annals}, 53(2):619--642, 2004.

\bibitem{athans1963}
Michael Athans.
\newblock Minimum-fuel feedback control systems: second-order case.
\newblock {\em IEEE Transactions on Applications and Industry}, 82(65):8--17,
  1963.

\bibitem{athans1964}
Michael Athans.
\newblock Fuel-optimal control of a double integral plant with response time
  constraints.
\newblock {\em IEEE Transactions on Applications and Industry},
  83(73):240--246, 1964.

\bibitem{Atherton75}
Derek Atherton.
\newblock {\em Nonlinear Control Engineering - Describing Function Analysis and
  Design}.
\newblock Workingam Beks, 1975.

\bibitem{aurich2009modelling}
Jan Aurich, Dirk Biermann, Heribert Blum, Christian Brecher, Carsten
  Carstensen, Berend Denkena, Fritz Klocke, Matthias Kr{\"o}ger, Paul
  Steinmann, and Klaus Weinert.
\newblock Modelling and simulation of process: machine interaction in grinding.
\newblock {\em Production Engineering}, 3:111--120, 2009.

\bibitem{bartolini1997}
Giorgio Bartolini, Antonella Ferrara, and Elio Usai.
\newblock Output tracking control of uncertain nonlinear second-order systems.
\newblock {\em Automatica}, 33(12):2203--2212, 1997.

\bibitem{bartolini1998}
Giorgio Bartolini, Antonella Ferrara, and Elio Usai.
\newblock Chattering avoidance by second-order sliding mode control.
\newblock {\em IEEE Transactions on Automatic Control}, 43(2):241--246, 1998.

\bibitem{bartolini2003}
Giorgio Bartolini, Alessandro Pisano, Elisabetta Punta, and Elio Usai.
\newblock A survey of applications of second-order sliding mode control to
  mechanical systems.
\newblock {\em International Journal of Control}, 76(9--10):875--892, 2003.

\bibitem{boiko2009}
Igor Boiko.
\newblock {\em Discontinuous control systems: frequency-domain analysis and
  design}.
\newblock Birkhaeuser, 2009.

\bibitem{boiko2006}
Igor Boiko, Leonid Fridman, Rafael Iriarte, Alessandro Pisano, and Elio Usai.
\newblock Parameter tuning of second-order sliding mode controllers for linear
  plants with dynamic actuators.
\newblock {\em Automatica}, 42(5):833--839, 2006.

\bibitem{boiko2007}
Igor Boiko, Leonid Fridman, Alessandro Pisano, and Elio Usai.
\newblock Analysis of chattering in systems with second-order sliding modes.
\newblock {\em IEEE Transactions on Automatic Control}, 52(11):2085--2102,
  2007.

\bibitem{brand2025active}
Ziv Brand, Matthew Cole, and Nikolay Razoronov.
\newblock An active tool holder and robust lpv control design for practical
  vibration suppression in internal turning.
\newblock {\em Control Engineering Practice}, 156:106215, 2025.

\bibitem{cumpson2004}
Peter Cumpson, Peter Zhdan, and John Hedley.
\newblock Calibration of {AFM} cantilever stiffness: a microfabricated array of
  reflective springs.
\newblock {\em Ultramicroscopy}, 100(3-4):241--251, 2004.

\bibitem{dai2018fast}
Gaoliang Dai, Ludger Koenders, Jens Fluegge, and Matthias Hemmleb.
\newblock Fast and accurate: high-speed metrological large-range {AFM} for
  surface and nanometrology.
\newblock {\em Measurement Science and Technology}, 29(5):054012, 2018.

\bibitem{edwards1998}
Christopher Edwards and Sarah Spurgeon.
\newblock {\em Sliding mode control: theory and applications}.
\newblock CRC Press, 1998.

\bibitem{elson1995}
Merle Elson and Jean Bennett.
\newblock Calculation of the power spectral density from surface profile data.
\newblock {\em Applied optics}, 34(1):201--208, 1995.

\bibitem{estrada2025}
Manuel Estrada, Michael Ruderman, Oscar Texis-Loaiza, and Leonid Fridman.
\newblock Hydraulic actuator control based on continuous higher order sliding
  modes.
\newblock {\em Control Engineering Practice}, 156:106218, 2025.

\bibitem{FridmanLevant2002}
Leonid Fridman and Arie Levant.
\newblock Higher order sliding modes.
\newblock In Wilfrid Peruquetti and Jean-Pierre Barbot, editors, {\em Sliding
  mode control in engineering}. Marcel Dekker, 2002.

\bibitem{fuller1960relay}
A.T. Fuller.
\newblock Relay control systems optimized for various performance criteria.
\newblock {\em IFAC Proceedings Volumes}, 1(1):520--529, 1960.

\bibitem{gong2016surface}
Yuxuan Gong, Scott Misture, Peng Gao, and Nathan Mellott.
\newblock Surface roughness measurements using power spectrum density analysis
  with enhanced spatial correlation length.
\newblock {\em The Journal of Physical Chemistry C}, 120(39):22358--22364,
  2016.

\bibitem{hrovat1997survey}
Davor Hrovat.
\newblock Survey of advanced suspension developments and related optimal
  control applications.
\newblock {\em Automatica}, 33(10):1781--1817, 1997.

\bibitem{ito2018}
Shingo Ito and Georg Schitter.
\newblock Atomic force microscopy capable of vibration isolation with
  low-stiffness z-axis actuation.
\newblock {\em Ultramicroscopy}, 186:9--17, 2018.

\bibitem{koch2020sliding}
Stefan Koch, Martin Kleindienst, Jaime Moreno, and Martin Horn.
\newblock Sliding mode control of a distributed-parameter wafer spin clean
  process.
\newblock {\em IEEE Transactions on Control Systems Technology},
  29(5):2271--2278, 2020.

\bibitem{LeeUtkin2007}
Hoon Lee and Vadim Utkin.
\newblock Chattering suppression methods in sliding mode control systems.
\newblock {\em Annual Reviews in Control}, 31(2):179--188, 2007.

\bibitem{levant1993}
Arie Levant.
\newblock Sliding order and sliding accuracy in sliding mode control.
\newblock {\em International journal of control}, 58(6):1247--1263, 1993.

\bibitem{liang2004machining}
Steven Liang, Rogelio Hecker, and Robert Landers.
\newblock Machining process monitoring and control: the state-of-the-art.
\newblock {\em J. of Manufacturing Science and Engineering}, 126(2):297--310,
  2004.

\bibitem{nayak1971}
Ranganath Nayak.
\newblock Random process model of rough surfaces.
\newblock {\em ASME Journal of Lubrication Technology}, 93(3):398--407, 1971.

\bibitem{persson2006}
Bo~Persson.
\newblock Contact mechanics for randomly rough surfaces.
\newblock {\em Surface science reports}, 61(4):201--227, 2006.

\bibitem{Pontryagin1962}
L.S. Pontryagin, V.G. Boltyanskii, R.V. Gamkrelidze, and E.~Mishchenko.
\newblock {\em The mathematical theory of optimal processes}.
\newblock Interscience Publishers, 1962.

\bibitem{rao2024adaptive}
Venkateswara Rao, Hamed Habibi, Jose~Luis Sanchez-Lopez, Prathyush Menon,
  Christopher Edwards, and Holger Voos.
\newblock Adaptive super-twisting controller design for accurate trajectory
  tracking performance of unmanned aerial vehicles.
\newblock {\em IEEE Transactions on Control Systems Technology},
  32(6):2126--2135, 2024.

\bibitem{rinaldi2021}
Gianmario Rinaldi, Prathyush Menon, Christopher Edwards, and Antonella Ferrara.
\newblock Sliding mode observer-based finite time control scheme for frequency
  regulation and economic dispatch in power grids.
\newblock {\em IEEE Tran. on Control Systems Technology}, 30(3):1296--1303,
  2021.

\bibitem{ruderman2023}
Michael Ruderman.
\newblock {\em Analysis and compensation of kinetic friction in robotic and
  mechatronic control systems}.
\newblock CRC Press, 2023.

\bibitem{ruderman2024}
Michael Ruderman.
\newblock Experimental benchmarking of energy-saving sub-optimal sliding mode
  control.
\newblock In {\em IEEE 17th International Workshop on Variable Structure
  Systems (VSS)}, pages 208--213, 2024.

\bibitem{ruderman2023energy}
Michael Ruderman, Alessandro Pisano, and Elio Usai.
\newblock Energy-saving sub-optimal sliding mode control with bounded
  actuation.
\newblock {\em arXiv preprint arXiv:2305.07891}, 2023.

\bibitem{shtessel2014}
Yuri Shtessel, Christopher Edwards, Leonid Fridman, and Arie Levant.
\newblock {\em Sliding mode control and observation}.
\newblock Springer, 2014.

\bibitem{sulchek2002}
Todd Sulchek, Goksen Yaralioglu, Calvin Quate, and Stephen Minne.
\newblock Characterization and optimization of scan speed for tapping-mode
  atomic force microscopy.
\newblock {\em Review of Scientific Instruments}, 73(8):2928--2936, 2002.

\bibitem{tavares2020}
Rafael Tavares, Michael Ruderman, Daan Menjoie, Joan Molina, and Miguel Dhaens.
\newblock Modeling and field-experiments identification of vertical dynamics of
  vehicle with active anti-roll bar.
\newblock In {\em IEEE 16th International Workshop on Advanced Motion Control},
  pages 161--167, 2020.

\bibitem{tseng2015}
Hongtei~Erik Tseng and Davor Hrovat.
\newblock State of the art survey: active and semi-active suspension control.
\newblock {\em Vehicle System Dynamics}, 53(7):1034--1062, 2015.

\bibitem{utkin1992}
Vadim Utkin.
\newblock {\em Sliding modes in control and optimization}.
\newblock Springer, 1992.

\bibitem{utkin2015}
Vadim Utkin.
\newblock Discussion aspects of high-order sliding mode control.
\newblock {\em IEEE Transactions on Automatic Control}, 61(3):829--833, 2015.

\bibitem{utkin2020}
Vadim Utkin, Alex Poznyak, Yury Orlov, and Andrey Polyakov.
\newblock {\em Road map for sliding mode control design}.
\newblock Springer, 2020.

\bibitem{wang2021advancing}
Ke~Wang, Kevin Taylor, and Lin Ma.
\newblock Advancing the application of atomic force microscopy {(AFM)} to the
  characterization and quantification of geological material properties.
\newblock {\em International Journal of Coal Geology}, 247:103852, 2021.

\bibitem{yong2012invited}
Yuen Yong, Reza Moheimani, Brian Kenton, and Kam Leang.
\newblock Invited review article: High-speed flexure-guided nanopositioning:
  Mechanical design and control issues.
\newblock {\em Review of scientific instruments}, 83(12), 2012.

\bibitem{yong2012design}
Yuen Yong and Reza Mohemani.
\newblock Design of an inertially counterbalanced {Z}-nanopositioner for
  high-speed atomic force microscopy.
\newblock {\em IEEE Trans. on Nanotechnology}, 12(2):137--145, 2012.

\bibitem{zuo2003}
Lei Zuo and Samir~A Nayfeh.
\newblock Structured {H2} optimization of vehicle suspensions based on
  multi-wheel models.
\newblock {\em Vehicle System Dynamics}, 40(5):351--371, 2003.

\end{thebibliography}

\end{document}